\documentclass[amsmath,amssymb,showkeys,superscriptaddress,aps,prd,floatfix,twocolumn,10pt]{revtex4-2}
\usepackage{graphicx}
\usepackage[utf8]{inputenc}
\usepackage[caption=false]{subfig}
\usepackage{multirow}
\usepackage{appendix}
\usepackage{graphicx,epstopdf}

\def\xslash{x\!\!\!\slash }

\def\vel{\left|}
\def\ver{\right|}
\usepackage{colortbl}
\usepackage{xcolor}
\usepackage[colorlinks=true, urlcolor=blue, linkcolor=blue, citecolor=green]{hyperref}

\begin{document}

\title{Magnetic dipole moments of  \texorpdfstring{$B_{(s)}^{(*)}B_{(s)}^{(*)}$}{}  states}
\author{Ula\c{s}~\"{O}zdem}%
\email[]{ulasozdem@aydin.edu.tr}
\affiliation{ Health Services Vocational School of Higher Education, Istanbul Aydin University, Sefakoy-Kucukcekmece, 34295 Istanbul, T\"{u}rkiye}

 
\begin{abstract}
We systematically study the magnetic dipole moments of multiquark states.
In this study, the magnetic dipole moments of possible $B^- B^{*-}$, $B^0 B^{*-}$, $B^- B^{*0} $, $B^0 B^{*0}$, $B_s^0 B^{*-}$, $B^- B_s^{*0}$, $B_s^{0} B^{*0}$, $B^0 B_s^{*0}$ and  $B^0_s B_s^{*0}$ states are extracted using light-cone sum rules.
We explore magnetic dipole moments of these states as molecular picture with spin-parity  $J^P = 1^+$. The magnetic dipole moments of hadrons include useful information on the distributions of internal charge and magnetization, which can be used to understand their geometrical shapes and quark-gluon organization. The results of the present study along with the spectroscopic parameters may help the future theoretical and experimental research on the characteristics of doubly-bottom tetraquark states.
\end{abstract}
\keywords{Magnetic dipole moment, doubly-bottom tetraquarks, $T_{bb}$ states, light-cone sum rules}

\maketitle

\section{Introduction}\label{introduction}

Recently, the LHCb Collaboration made a great breakthrough in the search for multiquark states and reported a new state ($ T_{cc}^{+}$ for short) below the $D^0  D^{\ast +}$ mass threshold in the $D^0 D^0 \pi^+$ invariant mass spectrum~\cite{LHCb:2021vvq,LHCb:2021auc}. The fact that this observed state contains two charm quarks and has an electrical charge makes it a good candidate for an exotic state with the quark content $cc\bar u \bar d$. 
The significance of the $T^+_{cc}$ state is the same as that of $X(3872$); therefore the newly discovered state provides an important new platform for both experimental and theoretical hadron physics. In literature, there are numerous  theoretical studies that have been performed to understand the spectroscopic parameters, magnetic dipole moments, production mechanisms and decay modes of doubly-charmed tetraquark states within different models~\cite{Qin:2020zlg,Feijoo:2021ppq,Deng:2021gnb,Dong:2021bvy,Yan:2021wdl,Wang:2021yld,Huang:2021urd,Agaev:2021vur,Chen:2021tnn,Jin:2021cxj,Ling:2021bir,Hu:2021gdg,Chen:2021cfl,Albaladejo:2021vln,Abreu:2021jwm,Du:2021zzh,Dai:2021vgf,Wang:2021ajy,Meng:2021jnw,Fleming:2021wmk,Xin:2021wcr,Ren:2021dsi,Albuquerque:2022weq, Azizi:2021aib,Ozdem:2021hmk,Kim:2022mpa,Abreu:2022lfy,Agaev:2022ast,Deng:2021gnb}. 

If  $T^+_{cc}$ is the doubly-charmed tetraquark state, there may also be doubly-bottom tetraquark states.  Even if these doubly-bottom tetraquark states do not exist, it is  important, in our opinion, to explore the reasons why. Inspired by this, we are well-motivated to search for the possible doubly-bottom tetraquark states. Therefore, besides the doubly-charmed states, the properties of doubly-bottom tetraquark states have also been extracted  in different configurations~\cite{Ren:2021dsi,Albuquerque:2022weq,Deng:2021gnb,Dai:2022ulk,Dai:2021wxi,Karliner:2017qjm,Eichten:2017ffp,Cheng:2020wxa,Braaten:2020nwp, Meng:2020knc,Dias:2011mi,Navarra:2007yw,Gao:2020ogo,Agaev:2020mqq,Agaev:2020zag,Agaev:2020dba,Agaev:2019lwh,Agaev:2018khe,Aliev:2021dgx,Mohanta:2020eed,Ke:2022vsi}. In Ref. \cite{Ren:2021dsi}, the mass and decay width of the $T_{cc}$ and $T_{bb}$ states were investigated in the framework of the one-boson exchange potential model. The researchers  predicted that $T_{bb}$ states are more stable than $T_{cc}$ states. In Ref. \cite{Dai:2022ulk}, the authors studied the interaction of  $T_{bb}$ states by means of vector meson exchange with Lagrangians from an extension of the local hidden-gauge approach. They predicted that only $B^*B$,  ($B_s^*B-B^*B_s$), $B^*B^*$ and $B_s^*B^*$ states form bound states with the quantum numbers $J^P =1^+$. In Ref.~\cite{Cheng:2020wxa}, the masses of the $QQ\bar q\bar q$ tetraquark states were obtained with the help of heavy diquark-antiquark symmetry and the chromomagnetic interaction model. They predicted that only the $bb \bar q \bar q$ and $bb \bar q \bar s$ states are stable with respect to  strong decays and discussed the constraints on the masses of the these tetraquark states. In Ref.~\cite{Braaten:2020nwp}, the authors predicted the masses of doubly-heavy tetraquark states in the heavy quark limit. They found that only doubly-bottom tetraquarks with  $\bar u \bar d$, $\bar s \bar u$ and $\bar s \bar d$ are stable with strong decays. In Ref. \cite{Meng:2020knc}, the bound states of doubly-heavy tetraquarks were investigated via the non-relativistic quark model. The authors obtained several stable states, one of which was a strongly bound $bb\bar q \bar q$ with isospin and spin-parity $I(J^P) =0 (1^+)$. In Ref. \cite{Aliev:2021dgx}, the spectroscopic parameters of $T_{bb}$ states were investigated within QCD sum rules using molecular pictures with quantum numbers $J^P =1^+$. Furthermore, in Ref. \cite{Mohanta:2020eed}, the authors attempted to extract possible $bb\bar u \bar d $ states within lattice QCD and found that one of these states form a bound state.  In addition to spectroscopic parameters, the semi leptonic and nonleptonic decays of double-bottom tetraquark states were extracted in the framework of the QCD sum rule method in Ref. \cite{Agaev:2020mqq, Agaev:2018khe}.

In Refs. \cite{Azizi:2021aib,Ozdem:2021hmk}, we extracted the magnetic dipole moments of doubly-charmed tetraquark states in the molecular picture using the light-cone  sum rule method.  For the case of $T^+_{cc}$, the magnetic dipole moment was also obtained by considering it as a diquark-antidiquark state. We extend our work to doubly-bottom tetraquark states. In this study, we evaluate the magnetic dipole moments of doubly-bottom tetraquark states in the molecular framework using the light-cone sum rule method \cite{Chernyak:1990ag, Braun:1988qv, Balitsky:1989ry}. The light-cone sum rules method has been employed in literature to obtain information about the dynamic and static parameters of conventional and non-conventional hadrons providing successful predictions that are consistent with experimental ones. The magnetic dipole moment of hadrons represents an important tool for understanding their internal structure in terms of quarks and gluons. Thus, it is important and also interesting  to investigate the magnetic dipole moments of conventional and non-conventional hadrons.

The paper is organized as follows: In Sec. \ref{formalism}, light-cone sum rules for the magnetic dipole moments of doubly-bottom tetraquark states are calculated. Sec. \ref{numerical} is devoted to the numerical calculations of the magnetic dipole moment sum rules and discussion. Explicit expressions for the magnetic dipole moments of the doubly-bottom tetraquark states and distribution amplitudes (DAs) of  photons are presented in the appendices.


 \section{Formalism} \label{formalism}
 
 In the light-cone sum rules method, the correlation function is evaluated  in terms of hadrons (hadronic side) and quark-gluon degrees of freedom (QCD side). Then, by applying the continuum subtraction and double Borel transformation to eliminate the effects from higher states and the continuum and enhance the contributions of the ground state, and matching these results, we can obtain the desired sum rules.

To identify the magnetic dipole moments of doubly-bottom  tetraquark states within the light-cone sum rules, we introduce the following correlation function: 
\begin{equation}
 \label{edmn01}
\Pi _{\mu \nu }(p,q)=i\int d^{4}xe^{ip\cdot x}\langle 0|\mathcal{T}\{J_{\mu}(x)
J_{\nu }^{ \dagger }(0)\}|0\rangle_{\gamma}, 
\end{equation}%
where   $\gamma$ indicates the external electromagnetic field and $J_{\mu}(x)$ is the interpolating current of $T_{bb}$ states with the spin-parity $ J^{P} = 1^{+}$. Following the molecular configuration for $T_{bb}$ states, we consider the interpolating current as
\begin{eqnarray}\label{curr}
   J_{\mu}(x)&= \big[\bar {q_1}^a(x) i\gamma_5 b^a(x)][\bar q_2^b(x) \gamma_\mu b^b(x)],
     \end{eqnarray}
where $q_1$ and $q_2$ denote the $u$, $d$ and $s$-quarks.


 To acquire the hadronic side of the correlation function we insert a complete set of hadronic states with quantum number $J^P =1^+$ into the correlation function and isolate the contributions of the ground state of doubly-bottom tetraquark states. As a result of these calculations, we obtain the following:
 \begin{align}
\label{edmn04}
\Pi_{\mu\nu}^{Had} (p,q) &= {\frac{\langle 0 \mid J_\mu (x) \mid
T_{bb}(p, \varepsilon^\theta) \rangle}{p^2 - m_{T_{bb}}^2}} \nonumber\\&\langle T_{bb}(p, \varepsilon^\theta) \mid T_{bb}(p+q, \varepsilon^\delta) \rangle_\gamma \nonumber\\
&
\frac{\langle T_{bb}(p+q,\varepsilon^\delta) \mid {J^\dagger}_\nu (0) \mid 0 \rangle}{(p+q)^2 - m_{T_{bb}}^2} \nonumber\\
&+\mbox{higher states}\,.
\end{align}

The amplitude $\langle 0 \mid J_\mu(x) \mid T_{bb}(p,\varepsilon^\theta) \rangle$ can be parameterized in terms of the residue $\lambda_{T_{bb}}$ and polarization vector $\varepsilon_\mu^{\theta}$  of  $ T_{bb}$ states as
\begin{align}
\label{edmn05}
\langle 0 \mid J_\mu(x) \mid T_{bb}(p,\varepsilon^\theta) \rangle &= \lambda_{T_{bb}} \varepsilon_\mu^\theta\,,
\end{align}
whereas  the matrix element $\langle T_{bb}(p,\varepsilon^\theta) \mid  T_{bb} (p+q,\varepsilon^{\delta})\rangle_\gamma$ is given by 
\begin{align}
\langle T_{bb}(p,\varepsilon^\theta) \mid  T_{bb} (p+q,\varepsilon^{\delta})\rangle_\gamma &= - \varepsilon^\tau (\varepsilon^{\theta})^\alpha (\varepsilon^{\delta})^\beta \bigg\{ G_1(Q^2)\nonumber\\
& \times (2p+q)_\tau ~g_{\alpha\beta}  + G_2(Q^2)\nonumber\\
& \times ( g_{\tau\beta}~ q_\alpha -  g_{\tau\alpha}~ q_\beta) \nonumber\\ &- \frac{1}{2 m_{T_{bb}}^2} G_3(Q^2)~ (2p+q)_\tau \nonumber\\
& \times q_\alpha q_\beta  \bigg\},\label{edmn06}
\end{align}
where $\varepsilon^\tau$ is the polarization of the photon,  and   $G_1(Q^2)$, $G_2(Q^2)$, and $G_3(Q^2)$ are electromagnetic form factors,  with  $Q^2=-q^2$.

Using Eqs. (\ref{edmn04})-(\ref{edmn06}) and after doing some necessary calculations, we obtain the hadronic side of the correlation function as follows:
\begin{align}
\label{edmn09}
 \Pi_{\mu\nu}^{Had}(p,q) &=  \frac{\varepsilon_\rho \, \lambda_{T_{bb}}^2}{ [m_{T_{bb}}^2 - (p+q)^2][m_{T_{bb}}^2 - p^2]}
 \Big\{ G_2 (Q^2) \nonumber\\
 & \times \Big(q_\mu g_{\rho\nu} - q_\nu g_{\rho\mu} -
\frac{p_\nu}{m_{T_{bb}}^2}  \big(q_\mu p_\rho - \frac{1}{2}
Q^2 g_{\mu\rho}\big) 
 + \nonumber\\
 &  +
\frac{(p+q)_\mu}{m_{T_{bb}}^2}  \big(q_\nu (p+q)_\rho+ \frac{1}{2}
Q^2 g_{\nu\rho}\big) \nonumber\\
&
-  
\frac{(p+q)_\mu p_\nu p_\rho}{m_{T_{bb}}^4} \, Q^2
\Big)
\nonumber\\
&
+\mbox{other independent structures}\Big\}.
\end{align}

To characterize the magnetic dipole moment, we demand the value of the $G_2(Q^2)$ form factor to be at $Q^2 = 0$. The magnetic form factor $F_M(Q^2)$ is determined as: 
\begin{align}
\label{edmn07}
&F_M(Q^2) = G_2(Q^2)\,,
\end{align}
 and the magnetic dipole  moment $\mu_{T_{bb}}$ is described in terms of $F_M(Q^2=0)$ as follows:
\begin{align}
\label{edmn08}
&\mu_{T_{bb}} = \frac{ e}{2\, m_{T_{bb}}} \,F_M(Q^2=0).
\end{align}

The next step in obtaining the analytical expressions of the magnetic dipole moment calculations is  used to calculate the QCD side of the correlation function,which can be obtained by inserting the expression of the interpolating current given in Eq. (\ref{curr}) into Eq. (\ref{edmn01}) and using the Wick theorem. As a result, we get 
\begin{eqnarray}
\Pi _{\mu \nu }^{\mathrm{QCD}}(p,q)&=&-i\int d^{4}xe^{ip\cdot x} \, \langle 0 \mid  \nonumber\\
&& \Big\{\mathrm{%
Tr}\Big[ \gamma _{5} S_{b}^{aa^{\prime }}(x)\gamma
_{5}S_{q_1}^{a^{\prime }a}(-x)\Big]    
\mathrm{Tr}\Big[ \gamma _{\mu }S_{b}^{bb^{\prime
}}(x) \nonumber\\
&& \times \gamma _{\nu }S_{q_2}^{bb^{\prime }}(-x)\Big] -\mathrm{Tr}\Big[ \gamma
_{5}S_{b}^{ab^{\prime }}(x) \gamma _{\nu}S_{q_2}^{b^{\prime }b}(-x)  \nonumber\\
&& \times \gamma _{\mu }S_{b}^{ba^{\prime }}(x)\gamma _{5}S_{q_1}^{a^{\prime }a}(-x)\Big]\Big\} \mid 0 \rangle_{\gamma} ,  \label{eq:QCDSide}
\end{eqnarray}%
 where $S_{q}(x)$ and $S_{b}(x)$ denote the light and bottom-quark propagators.  
During our calculations, we utilize the x-space expressions for the light and bottom-quark propagators,
\begin{align}
\label{edmn12}
S_{q}(x)&=i \frac{{\xslash}}{2\pi ^{2}x^{4}} 
- \frac{\langle \bar qq \rangle }{12} \Big(1-i\frac{m_{q} \xslash}{4}   \Big)
- \frac{ \langle \bar qq \rangle }{192}m_0^2 x^2   \nonumber\\
& \times \Big(1-i\frac{m_{q} \xslash}{6}   \Big)
-\frac {i g_s }{32 \pi^2 x^2} ~G^{\mu \nu} (x) \Big[\rlap/{x} 
\sigma_{\mu \nu} +  \sigma_{\mu \nu} \rlap/{x}
 \Big],
\end{align}
\begin{align}
\label{edmn13}
S_{b}(x)&=\frac{m_{b}^{2}}{4 \pi^{2}} \Bigg[ \frac{K_{1}\Big(m_{b}\sqrt{-x^{2}}\Big) }{\sqrt{-x^{2}}}
+i\frac{{\xslash}~K_{2}\Big( m_{b}\sqrt{-x^{2}}\Big)}
{(\sqrt{-x^{2}})^{2}}\Bigg] \nonumber\\
&
-\frac{g_{s}m_{b}}{16\pi ^{2}} \int_0^1 dv\, G^{\mu \nu }(vx)\Bigg[ \big(\sigma _{\mu \nu }{\xslash}
  +{\xslash}\sigma _{\mu \nu }\big)\nonumber\\
  &\times \frac{K_{1}\Big( m_{b}\sqrt{-x^{2}}\Big) }{\sqrt{-x^{2}}}
+2\sigma_{\mu \nu }K_{0}\Big( m_{b}\sqrt{-x^{2}}\Big)\Bigg],
\end{align}%
where $\langle \bar qq \rangle$ is the light-quark  condensate, $m_0$ is characterized via the quark-gluon mixed condensate  $\langle 0 \mid \bar  q\, g_s\, \sigma_{\alpha\beta}\, G^{\alpha\beta}\, q \mid 0 \rangle = m_0^2 \,\langle \bar qq \rangle $, $v$ is the line variable, $G^{\mu\nu}$ is the gluon field strength tensor, and $K_1$, $K_2$, and $K_3$ are modified Bessel functions of the second kind. 
%


\begin{figure}[htp]\label{Feyndiag}
\subfloat[]{ \includegraphics[width=0.47\textwidth]{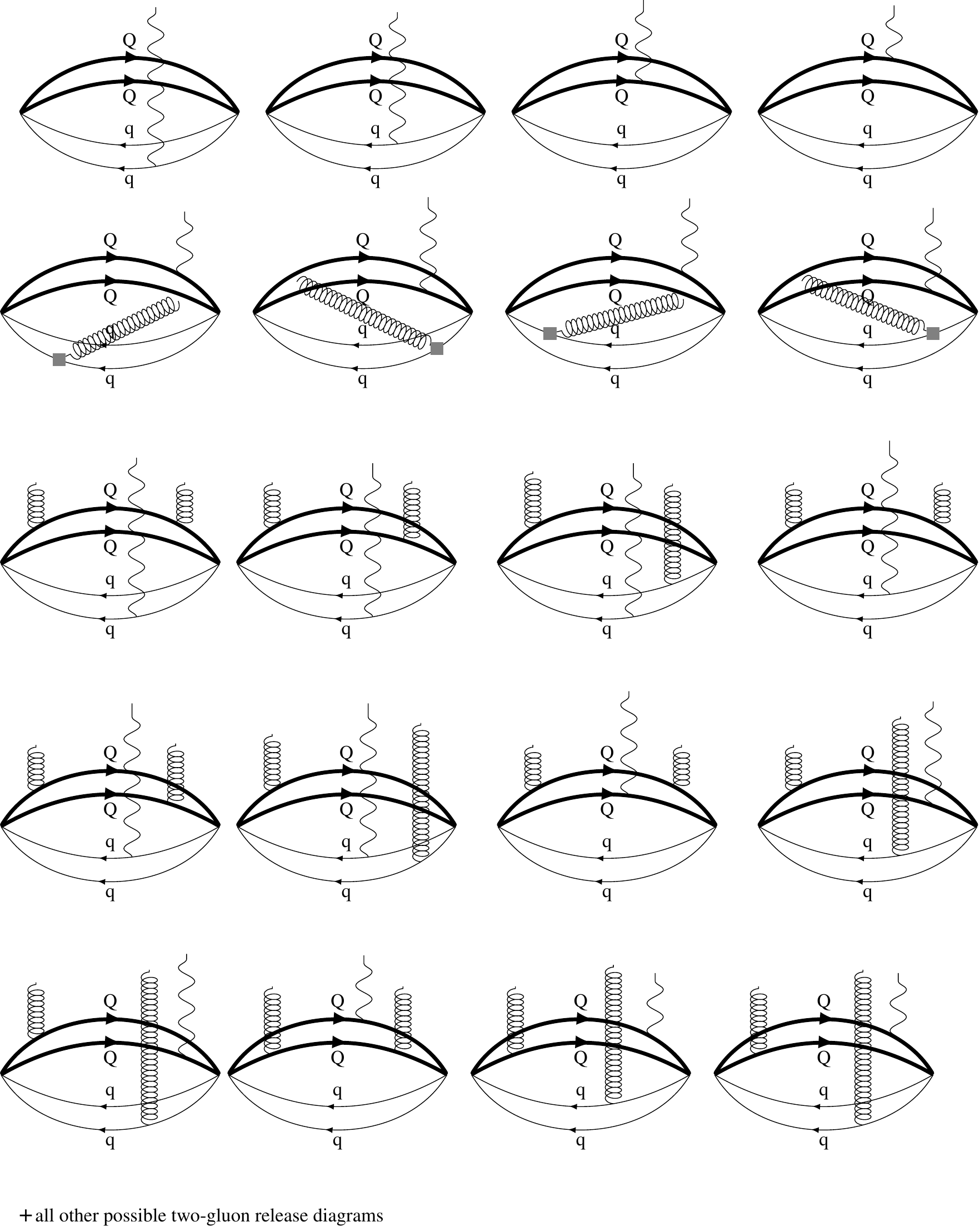}}\\
\subfloat[]{ \includegraphics[width=0.47\textwidth]{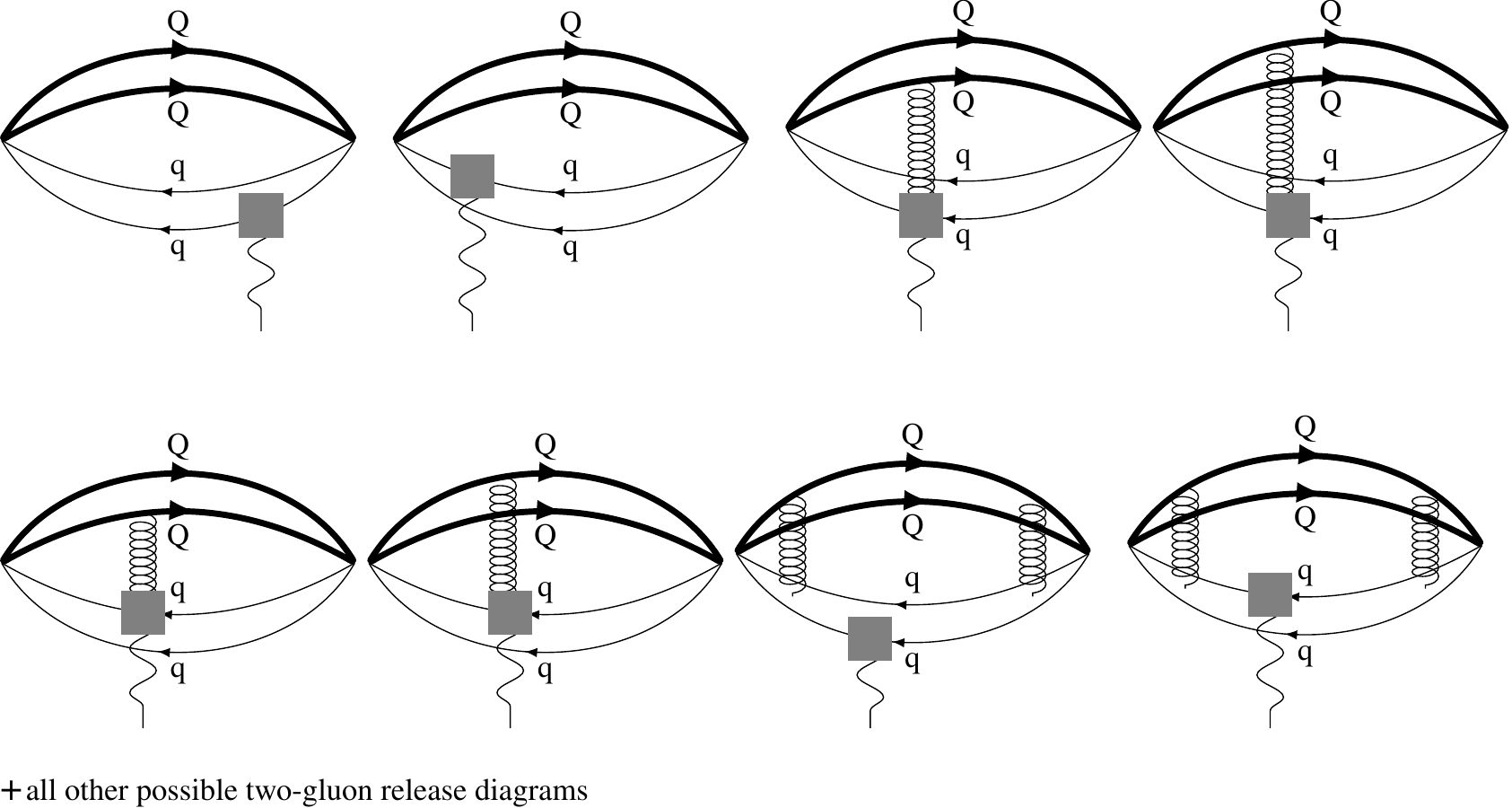}}
 \caption{ Feynman diagrams of the  magnetic dipole moments of doubly-bottom tetraquark states. The thick, thin, wavy and
curly lines represent the heavy quark, light quark, photon and gluon propagators, respectively. 
 Diagrams (a) correspond to the perturbative photon vertex and,  
 diagrams (b) represent the contributions originating from the DAs of the photon.}
  \end{figure}
  
In Fig. 1,  we present some of the possible Feynman diagrams that contribute to the QCD side of the analysis. For simplicity, we do not presented the Feynman diagrams of terms proportional  to  higher-dimensional operators, but they are considered in numerical analysis.

The correlation function in Eq. (\ref{eq:QCDSide}) includes different contributions,that is, the photon can be emitted both perturbatively (contributions of short-distance) and non-perturbatively (contributions of long-distance). In the first case, the photon perturbatively interacts with one of the light or heavy quarks; To obtain this contribution,  
the propagator of the quark interacting with the photon perturbatively is modified via
\begin{align}
\label{free}
S^{free}(x) \rightarrow \int d^4y\, S^{free} (x-y)\,\rlap/{\!A}(y)\, S^{free} (y)\,,
\end{align}
where $S^{free}(x)$ is the first term of the light and bottom quark propagators, and  the remaining three propagators in Eq.~(\ref{eq:QCDSide}) are replaced with full quark propagators involving the perturbative and non-perturbative contributions. In the second case, one of the light quark propagators in Eq.~(\ref{eq:QCDSide}), defined as the photon emission at large distances, is replaced via
\begin{align}
\label{edmn14}
S_{\mu\nu}^{ab}(x) \rightarrow -\frac{1}{4} \big[\bar{q}^a(x) \Gamma_i q^b(x)\big]\big(\Gamma_i\big)_{\mu\nu},
\end{align}
 where  $S_{\mu\nu}^{ab}(x)$ is one of the light quark propagators given in Eq. (\ref{edmn12}),  $\Gamma_i = I, \gamma_5, \gamma_\mu, i\gamma_5 \gamma_\mu, \sigma_{\mu\nu}/2$,
and the remaining light and heavy quark propagators are replaced with the full quark propagators. Because a photon interacts with light-quark fields at long distances, there are the matrix elements of nonlocal operators between the vacuum and photon state such as $\langle \gamma(q)\vel \bar{q}(x) \Gamma_i q(0) \ver 0\rangle$ and $\langle \gamma(q)\vel \bar{q}(x) \Gamma_i G_{\mu\nu}q(0) \ver 0\rangle$.
These matrix elements are described with respect to the DAs of the photon, which are defined in Ref. \cite{Ball:2002ps}. After these calculations, the QCD side of the correlation function is obtained.

In the final step, by applying a double Borel transformation over the variables $-p^2$ and $-(p + q)^2$, choosing the coefficients of the same Lorentz structures ($\varepsilon.p \,(p_\mu q_\nu-p_\nu q_\mu)$) in both the  QCD and hadronic sides and equating them, and performing  quark-hadron duality approximation, we obtain the required light-cone sum rules for these magnetic dipole moments:
\begin{align}
 &\mu_{T_{bb}} = m_{T_{bb}}^2\, \frac{e^{\frac{m_{T_{bb}}^2}{M^2}}}{\lambda_{T_{bb}}^2} \,\, \Delta (M^2,s_0).
\end{align}

The explicit expression of the $\Delta (M^2,s_0)$ function is given in Appendix A.


\section{Numerical analysis and conclusions}\label{numerical}

In this section, numerical computations are performed for the magnetic dipole moments of  doubly-bottom tetraquark states.
The light-cone sum rules contain various input parameters, such as the light and heavy quark masses; and  light-quark and gluon condensates.    These parameters are given as: $m_u=m_d=0$, $m_s =96^{+8}_{-4}\,\mbox{MeV}$,
$m_b = (4.78\pm 0.06)\,$GeV, 
$f_{3\gamma}=-0.0039~GeV^2$~\cite{Ball:2002ps},  
$\langle \bar ss\rangle $= $0.8 \langle \bar uu\rangle$ with 
$\langle \bar uu\rangle $=$(-0.24\pm0.01)^3\,$GeV$^3$~\cite{Ioffe:2005ym},  
$m_0^{2} = 0.8 \pm 0.1$~GeV$^2$~\cite{Ioffe:2005ym}, $\langle g_s^2G^2\rangle = 0.88~ $GeV$^4$~\cite{Matheus:2006xi} and  $\chi=-2.85 \pm 0.5$~GeV$^{-2}$~\cite{Rohrwild:2007yt}. 
To obtain a numerical value for the magnetic dipole moments, we must define the values of the mass and residue of the doubly-bottom tetraquark states. These parameters are borrowed from Ref.~\cite{Aliev:2021dgx}. Another set of crucial input parameters are the photon DAs of different twists. Explicit expression of  these DAs is given in Appendix B.

It follows from the explicit expressions of the light-cone sum rules for the magnetic dipole moments of the doubly-bottom tetraquark states that; in addition to DAs, they contain two arbitrary parameters, namely, the Borel mass parameter $M^2$ and continuum threshold $s_0$. According to the light-cone sum rules methodology, we must find working intervals of these arbitrary parameters, where the magnetic dipole moments are insensitive to  variations in these arbitrary parameters in their working intervals. However, in practice, it is necessary to find the working region where the variation in the calculations of the magnetic dipole moments of  doubly-bottom tetraquark states according to these parameters is minimum. The continuum threshold $s_0$ is not arbitrary and is related to the energy of the first excited state in the initial channel. However, because we have very limited information on the energy of excited states, we must decide on the method of choosing the working interval of $s_0$. There are various proposals on how to determine this parameter. The  analysis of various sum rules predicts that  $s_0 \simeq (m_{ground}+0.5^{+0.1}_{-0.1})^2$~GeV$^{2}$.  
We use two criteria to determine the working region of $M^2$. The lower bound of $M^2$ is constrained by  operator product expansion (OPE) convergence, demanding the higher twist and higher condensate terms to less than $10 \%$ of the total. The upper bound of $M^2$ is constrained by the pole contribution (PC)
\begin{align}
 \mbox{PC} =\frac{\Delta (M^2,s_0)}{\Delta (M^2,\infty)},
 \end{align}
which represents the lowest-lying state contribution to the correlation function.
Our numerical calculations indicate that the requirements of the light-cone sum rules method are satisfied in the working intervals of the arbitrary parameters presented in Table \ref{parameter}. In our computations, for the magnetic dipole moments of doubly-bottom tetraquark states, the PC varies on average within the limits $0.35 - 0.61$. In the standard analysis of sum rules, the PC is expected to be larger than $0.5$ for conventional hadrons. However, in the case of tetraquark states, PC $> 0.2$.
When we analyze the OPE convergence, we have obtained that the contribution of the higher dimensional term in OPE is less than $\sim 1 \%$. 
As these results suggest, the chosen working intervals for $M^2$ and $s_0$ satisfy the requirements of the light-cone sum rules method. In Fig. 2, we present the dependence of the magnetic dipole moments of doubly-bottom tetraquark states on $M^2$ at various $s_0$ values. As shown, the variation in  magnetic dipole moment with respect to $M^2$ is roughly $10-15\%$. Although the magnetic dipole moments indicate some dependence on $s_0$, it remains inside the limits allowed by the light-cone sum rules and generates most of the uncertainties.
\begin{table}[t]
	\addtolength{\tabcolsep}{7pt}
	\caption{Working intervals of  $s_0$, $M^2$ and the PC for magnetic dipole moments.}
	\label{parameter}
		\begin{center}
\begin{tabular}{l|c|c|ccc}
	   \hline\hline
   $T_{bb}$ States&   $s_0$~\mbox{[GeV$^2$]}&$M^2$~\mbox{[GeV$^2$]}& PC\\
\hline\hline
$B^- B^{*-}$        &$ 115-119$        &  $ 11-15 $&$0.38-0.61$\\
$B^0 B^{*-}$        &$ 115-119$        &  $ 11-15 $&$0.35-0.58$\\
$B^- B^{*0} $       &$ 115-119$        &  $ 11-15 $&$0.35-0.61$\\
$B^0 B^{*0}$        &$ 115-119$        &  $ 11-15 $&$0.37-0.59$\\
$B_s^0 B^{*-}$      &$ 117-121$        &  $ 11-15 $&$0.36-0.58$\\
$B^- B_s^{*0}$      &$ 117-121$        &  $ 11-15 $&$0.37-0.60$\\
$B_s^{0} B^{*0}$    &$ 117-121$        &  $ 11-15 $&$0.36-0.59$\\
$B^0 B_s^{*0}$      &$ 117-121$        &  $ 11-15 $&$0.35-0.61$\\
$B^0_s B_s^{*0}$    &$ 121-125$        &  $ 11-15 $&$0.35-0.60$\\
	   \hline\hline
\end{tabular}
\end{center}
\end{table}

After performing the numerical calculations, the acquired values of the magnetic dipole moments of doubly-bottom tetraquark states are collected in Table \ref{MDMres}. The uncertainties on the results are due to the variation in the $s_0$, $M^2$ and, errors in the values of the input parameters. 
\begin{table}[t]
	\addtolength{\tabcolsep}{10pt}
	\caption{Numerical values of the magnetic dipole moments (in units of nuclear magneton $\mu_N$).}
	\label{MDMres}
		\begin{center}
\begin{tabular}{lccccc}
	   \hline\hline
   $T_{bb}$ States&   \mbox{Magnetic dipole moment}\\
\hline\hline
\\
$B^- B^{*-}$        &$ ~1.72 \pm 0.67$        \\
\hline
\\ 
$B^0 B^{*-}$        &$ ~1.38 \pm 0.56$        \\
\hline
\\
$B^- B^{*0} $       &$ -0.44 \pm 0.17$        \\
\hline
\\
$B^0 B^{*0}$        &$ -0.77 \pm 0.28$        \\
\hline
\\
$B_s^0 B^{*-}$      &$ ~1.47 \pm 0.42$        \\
\hline
\\
$B^- B_s^{*0}$      &$ -0.42 \pm 0.15$        \\
\hline
\\
$B_s^{0} B^{*0}$    &$ -0.77 \pm 0.25$        \\
\hline
\\
$B^0 B_s^{*0}$      &$ -0.73 \pm 0.24$        \\
\hline
\\
$B^0_s B_s^{*0}$    &$ -1.11 \pm 0.31$        \\
	   \hline\hline
\end{tabular}
\end{center}
\end{table}
The magnitude of the magnetic dipole moments indicates their measurability in experiments. From this perspective, it can be said that it is possible to measure the magnetic dipole moments acquired experimentally. $SU(3)_f$ breaking effects are considered through a nonzero s-quark mass and s-quark condensate, and  we predict that $SU(3)_f$ symmetry violation in the magnetic dipole moments is small, except for the relation between $B^0 B^{*0}$ and $B_sB_s^*$, where  $SU(3)_f$ symmetry violation is  large.  In Ref. \cite{Deng:2021gnb}, the authors  performed a systematic investigation of the magnetic dipole moments of doubly heavy tetraquark states with the molecule configuration within the framework of the non-relativistic quark model with the help of the Gaussian expansion method. The obtained magnetic dipole moments depending on their spatial configurations are given as $\mu_{T_{bb}^-}=0.49-0.98~ \mu_N$ and $\mu_{T_{bbs}^-}=1.29-1.40~ \mu_N$.

As shown by the analytical results given in the appendix, we choose to keep the quark charge factors explicit. The advantage of this is that it can make it possible to investigate individual quark contribution to the magnetic dipole moment. 
Via a profound analysis, it is observed that the sign of the magnetic dipole moment is determined by the $q_2$-quark in the interpolating current.  If $q_2$ is a $u$-quark, the sign of the magnetic dipole moment is positive ($B^- B^{*-}$, $B^0 B^{*-}$ and $B_s^0 B^{*-}$), and if  $q_2$ is one of the $d$ or $s$-quarks, the sign of the magnetic dipole moment is negative ($B^- B^{*0} $, $B^0 B^{*0}$, $B^- B_s^{*0}$, $B_s^{0} B^{*0}$, $B^0 B_s^{*0}$, and  $B^0_s B_s^{*0}$). A detailed examination indicates that the smallness of the $e_b$ and $e_{q_1}$ contributions are due to an almost exact cancellation of  terms involving  $e_b$ and $e_{q_1}$.  We  would also like to discuss the amount of  perturbative and non-perturbative contributions (for example,quark-gluon condensates; and the DAs of the photon) to the total results. Our numerical computations indicate that roughly $85\%$ of the total contribution belongs to the perturbative part, and the surviving $15\%$ corresponds to the non-perturbative contributions.

In conclusion, we extracted the magnetic dipole moments within the light-cone sum rules method by employing molecular type interpolating currents for  doubly bottom tetraquark states with spin-parity $J^P = 1^+$. The acquired results along with the spectroscopic parameters may help the future theoretical and experimental research on the characteristics of doubly-bottom tetraquark states. It
would be exciting to predict future experimental attempts at exploring possible doubly-bottom tetraquark states and test the results of the present analysis.

\section{Acknowledgements}

We are grateful to A. Ozpineci for useful discussions, comments and remarks.

\begin{widetext}
 
 \begin{figure}[htp]
\centering
\subfloat[]{\includegraphics[width=0.33\textwidth]{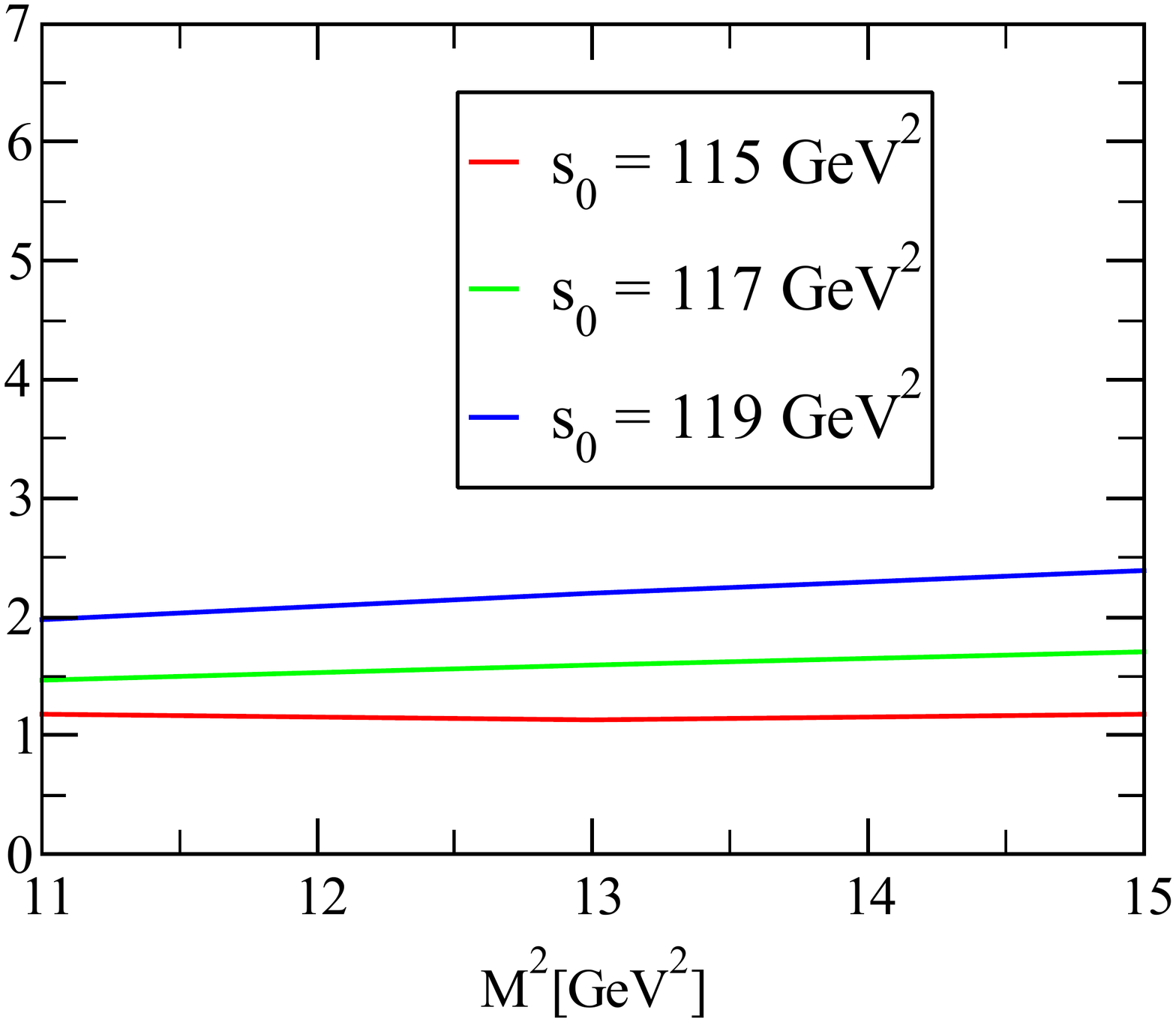}}
\subfloat[]{\includegraphics[width=0.33\textwidth]{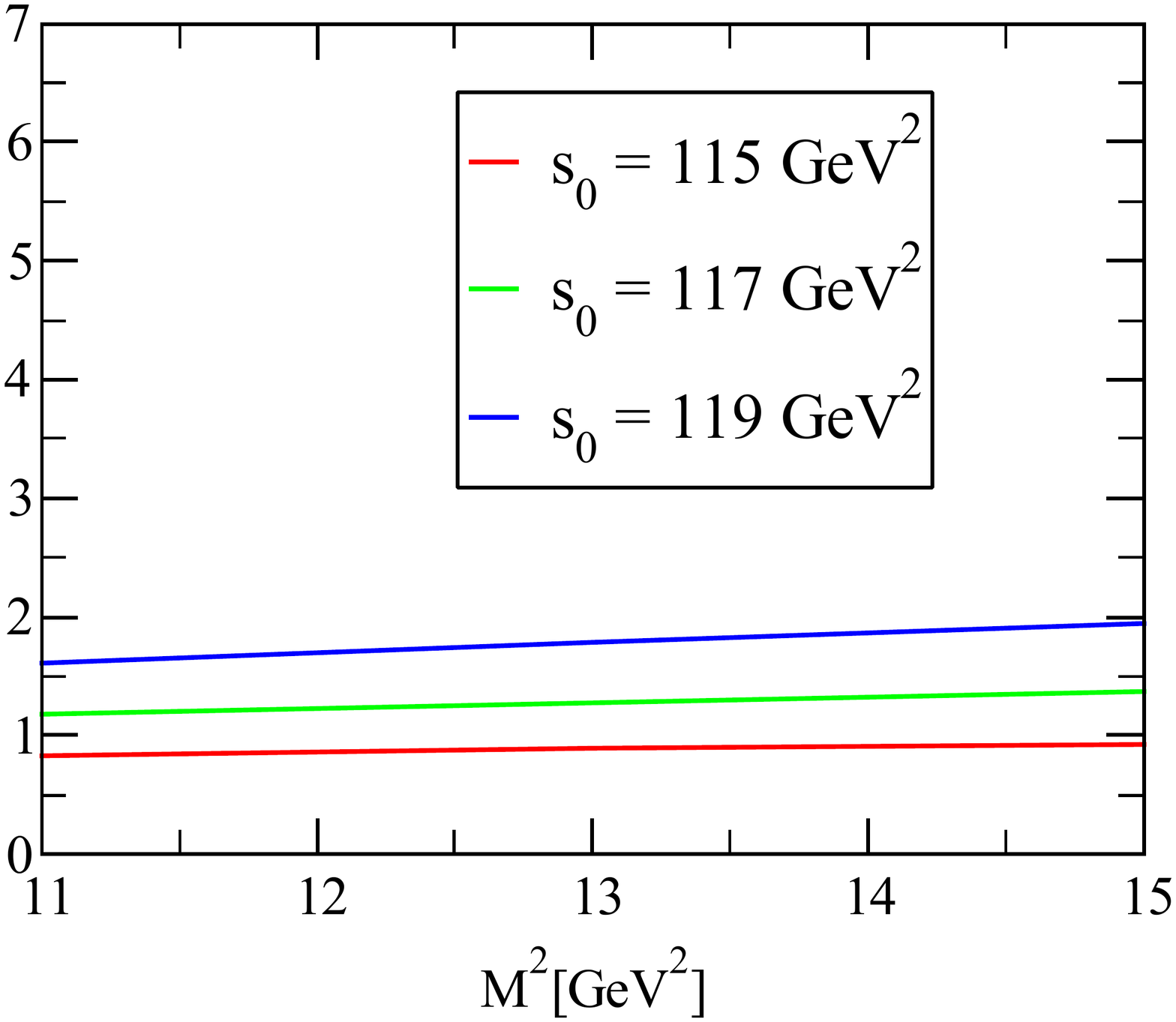}}
\subfloat[]{\includegraphics[width=0.33\textwidth]{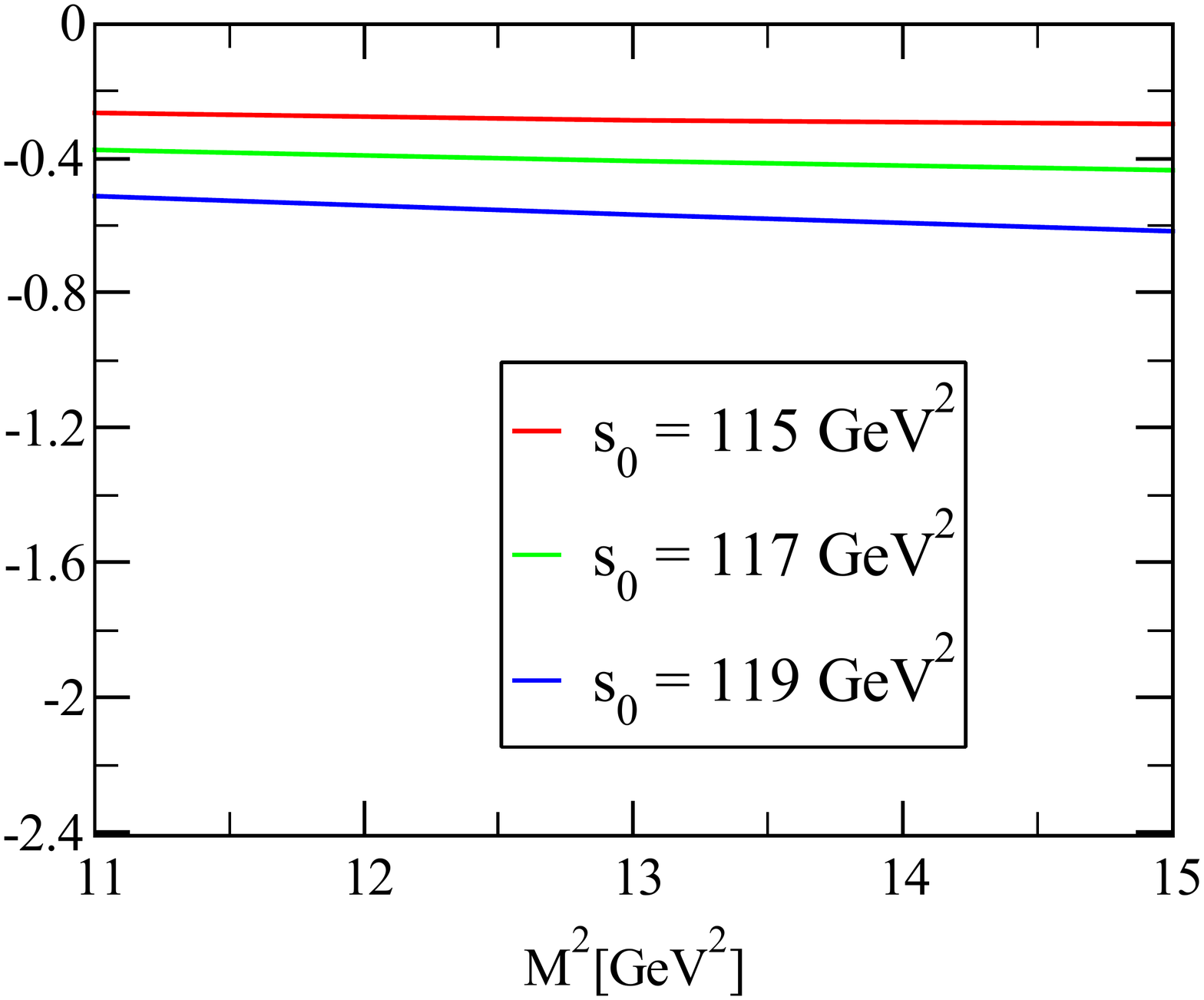}}\\ 
\subfloat[]{\includegraphics[width=0.33\textwidth]{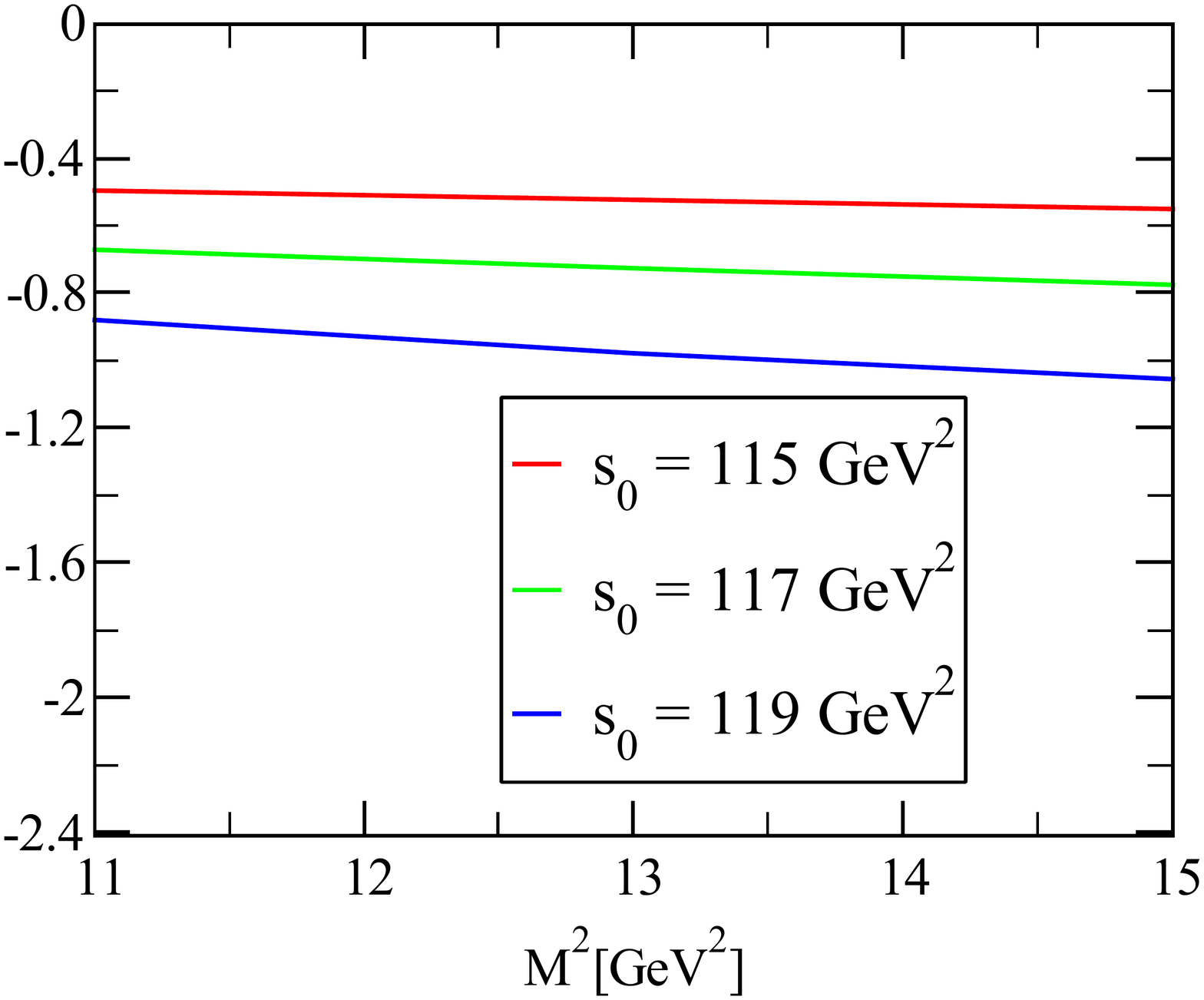}}
\subfloat[]{\includegraphics[width=0.33\textwidth]{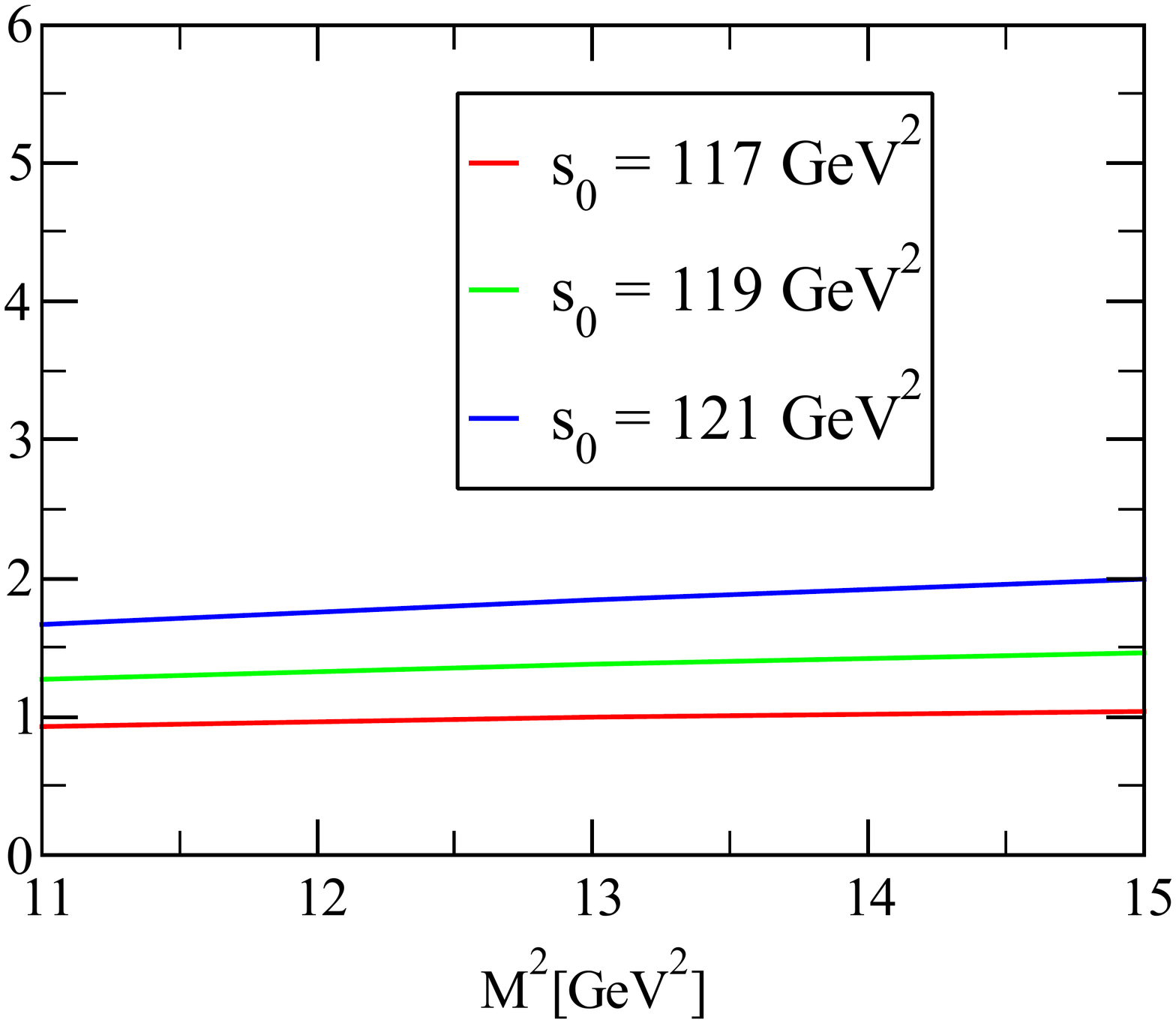}}
\subfloat[]{\includegraphics[width=0.33\textwidth]{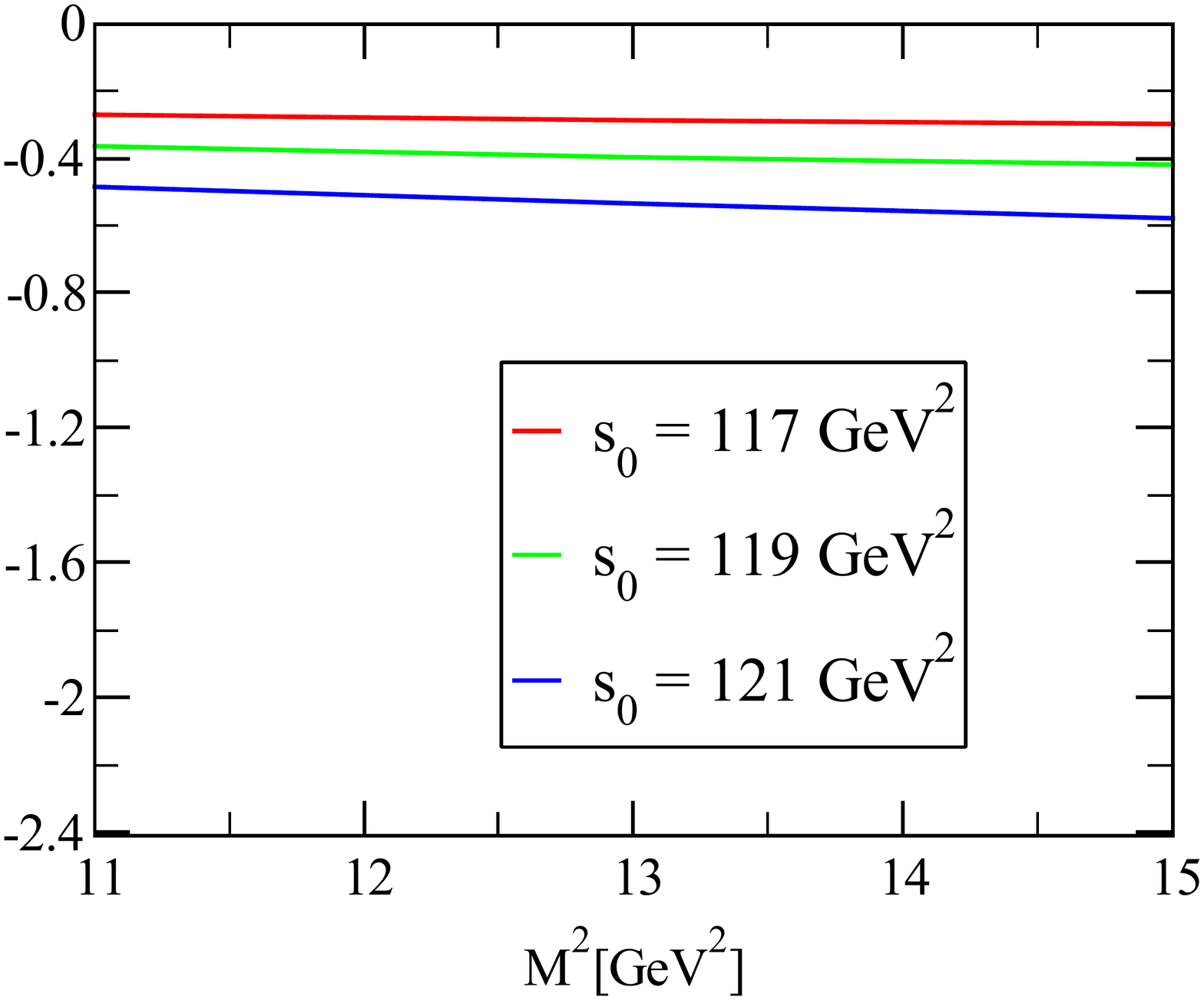}}\\
\subfloat[]{\includegraphics[width=0.33\textwidth]{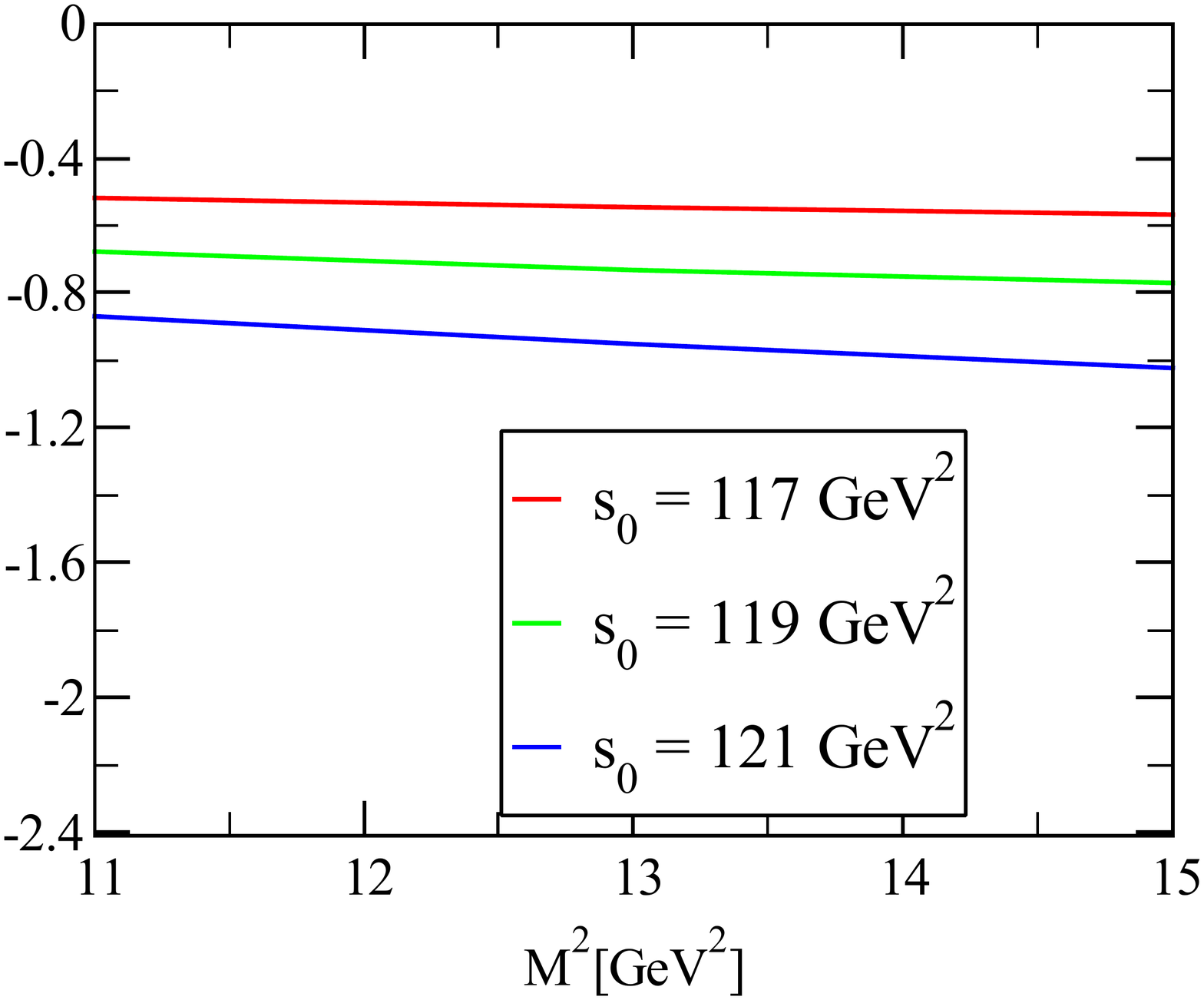}}
\subfloat[]{\includegraphics[width=0.33\textwidth]{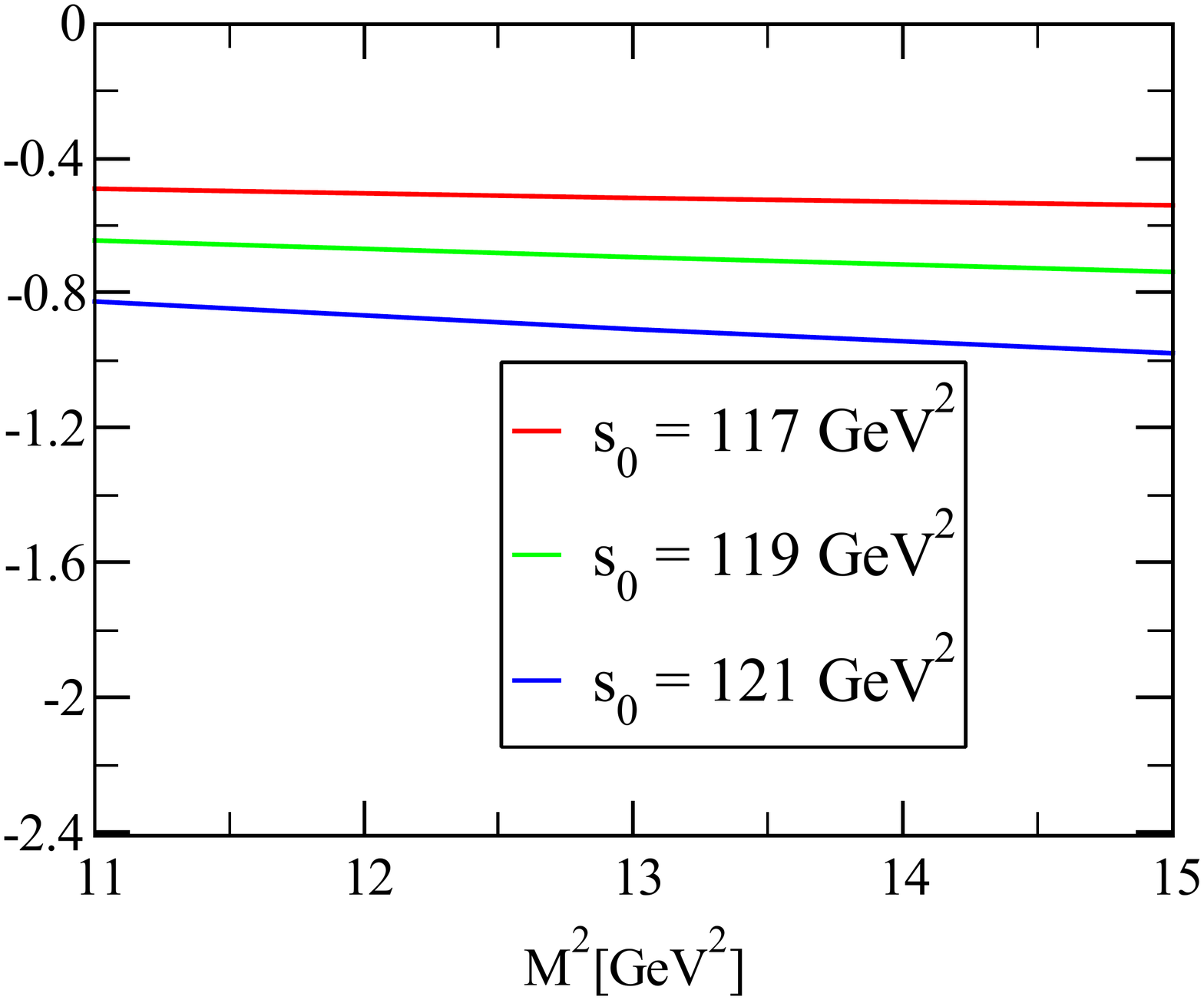}}
\subfloat[]{\includegraphics[width=0.33\textwidth]{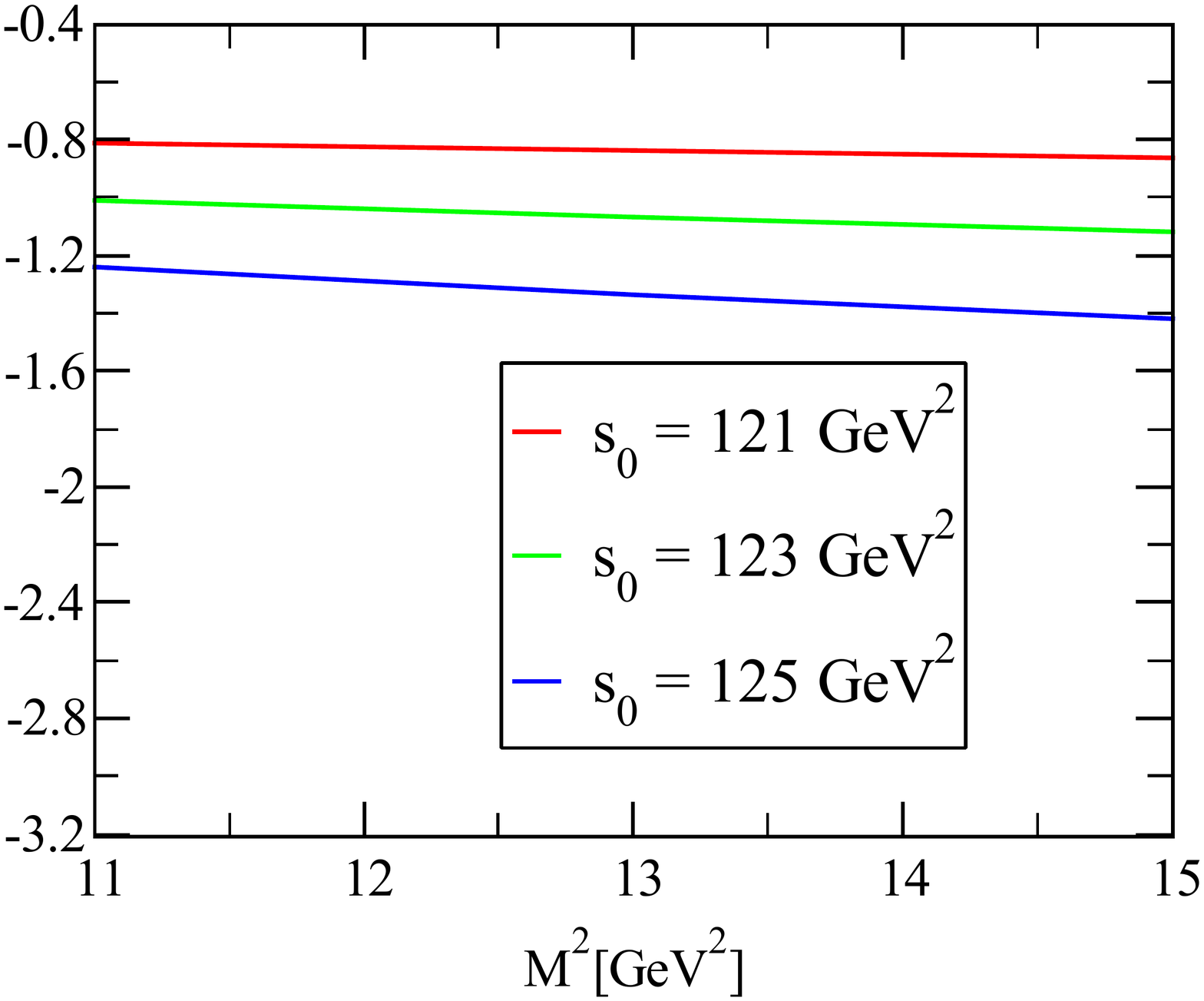}}\\
 \caption{Dependence of the magnetic dipole moments of doubly-bottom tetraquark states on $M^2$ at three different values of $s_0$; (a), (b), (c), (d), (e),  (f), (g), (h) and (i) represent $B^- B^{*-}$, $B^0 B^{*-}$, $B^- B^{*0} $, $B^0 B^{*0}$, $B_s^0 B^{*-}$, $B^- B_s^{*0}$, $B_s^{0} B^{*0}$, $B^0 B_s^{*0}$, and  $B^0_s B_s^{*0}$ states, respectively. }
 \label{Msqfig}
  \end{figure}
  
  \end{widetext}

  \begin{widetext}
  \appendix
  \subsection*{Appendix  A: Explicit expression for \texorpdfstring{$\Delta (M^2,s_0)$}{}}
 In Appendix A, we present the explicit expressions of the function $\Delta (M^2,s_0)$ for the magnetic dipole moments of  doubly-bottom tetraquark states entering the sum rule.
  \begin{align}
   \Delta (M^2,s_0) &= \frac {e_{q_1} m_b \langle g_s^2 G^2 \rangle \langle \bar q_1 q_1 \rangle} {3538944 \pi^3} \Bigg[ 
   8  \Big(I[0, 1, 1, 0] - 2 I[0, 1, 1, 1] + I[0, 1, 1, 2] - 
        2 I[0, 1, 2, 0] + 2 I[0, 1, 2, 1]\Big)\mathcal{A}[
      u_ 0] \nonumber\\
      &+ (8 A[u_0] + 3 I_ 3[\mathcal{S}] - 
       8 I_3[\mathcal{\tilde S}]) I[0, 1, 3, 0] - 
    4 \chi \Big(9 I[0, 2, 1, 0] - 23 I[0, 2, 1, 1] + 
        19 I[0, 2, 1, 2] - 5 I[0, 2, 1, 3]
        \nonumber\\
      &- 26 I[0, 2, 2, 0] + 
        44 I[0, 2, 2, 1] - 18 I[0, 2, 2, 2] + 25 I[0, 2, 3, 0] - 
        21 I[0, 2, 3, 1] - 8 I[0, 2, 4, 0]\Big)\varphi_{\gamma}[
       u_0]\Bigg] 
       \nonumber\\
       &-\frac {m_b \langle g_s^2 G^2\rangle \langle \bar q_2 q_2 \rangle } {3538944 \pi^3} \Bigg[-13 e_{q_ 2} \big (I_3[\mathcal {S}] + I_3[\mathcal {\tilde S}]\big) I[0, 1, 3, 0] + 
    24 e_b \Big (I[0, 1, 1, 0] - 2 I[0, 1, 1, 1] + I[0, 1, 1, 2] 
    \nonumber\\
       &- 
        2 I[0, 1, 2, 0] + 2 I[0, 1, 2, 1] + I[0, 1, 3, 0]\Big)\Bigg]
        \nonumber\\
        &
        -\frac {\langle g_s^2 G^2\rangle f_{3 \gamma}} {28311552 \pi^3}\Bigg[
   9 e_{q_ 1} I_1[\mathcal{V}] I[0, 2, 4, 0] + 
    48 e_{q_2} \Big (4 m_b (m_b + 2 m_{q_1}) \big (I[0, 1, 1, 
           0] - 2 I[0, 1, 1, 1] + I[0, 1, 1, 2] 
           \nonumber\\
        &- 2 I[0, 1, 2, 0] + 
          2 I[0, 1, 2, 1] + I[0, 1, 3, 0]\big) + I[0, 2, 2, 0] - 
       2 I[0, 2, 2, 1] + I[0, 2, 2, 2] - 2 I[0, 2, 3, 0]
       \nonumber\\
        &+ 
       2 I[0, 2, 3, 1] + I[0, 2, 4, 0]\Big) + 
    16 e_{q_1} \Big (4 m_b (m_b + 2 m_{q_2}) \big (I[0, 1, 1, 
            0] - 2 I[0, 1, 1, 1] + I[0, 1, 1, 2] 
            \nonumber\\
        &- 2 I[0, 1, 2, 0] + 
           2 I[0, 1, 2, 1] + I[0, 1, 3, 0]\big) + I[0, 2, 2, 0] - 
        2 I[0, 2, 2, 1] + I[0, 2, 2, 2] - 2 I[0, 2, 3, 0] 
        \nonumber\\
        &+ 
        2 I[0, 2, 3, 1] + I[0, 2, 4, 0]\Big) \psi^a[u_0]\Bigg]
        \nonumber\\
        &
        -\frac {\langle g_s^2 G^2\rangle} {28311552 \pi^3}\Bigg[
   216 e_ {q_ 2} \Big (I[0, 3, 2, 0] - 3 I[0, 3, 2, 1] + 
       3 I[0, 3, 2, 2] - I[0, 3, 2, 3] - 3 I[0, 3, 3, 0] - 
       6 I[0, 3, 3, 1]
       \nonumber\\
        &+ 3 I[0, 3, 3, 2] - 3 I[0, 3, 4, 0] + 
       3 I[0, 3, 4, 1] - I[0, 3, 5, 0]\Big) + 
    e_b \Bigg (-12 m_b \Big (6 m_ {q_ 2} \big (I[0, 2, 1, 1] - 
              2 I[0, 2, 1, 2]
              \nonumber\\
        &+ I[0, 2, 1, 3] - 2 I[0, 2, 2, 1] + 
              2 I[0, 2, 2, 2] + I[0, 2, 3, 1]\big) + 
           m_b \big (I[0, 2, 1, 1] - 10 I[0, 2, 1, 2] + 
               9 I[0, 2, 1, 3] 
               \nonumber\\
        &- 2 I[0, 2, 2, 1] + 10 I[0, 2, 2, 2] + 
               I[0, 2, 3, 1]\big)\Big) - 3 I[0, 3, 2, 0] - 
        127 I[0, 3, 2, 1] + 263 I[0, 3, 2, 2] 
        \nonumber\\
        &- 133 I[0, 3, 2, 3] + 
        9 I[0, 3, 3, 0] + 254 I[0, 3, 3, 1] - 263 I[0, 3, 3, 2] - 
        9 I[0, 3, 4, 0] - 127 I[0, 3, 4, 1] \nonumber\\
        & + 
        3 I[0, 3, 5, 0]\Bigg)\Bigg] \nonumber\\
        &
        +\frac {m_b \langle \bar q_1 q_1 \rangle} {196608 \pi^3}\Bigg[
   72 e_ {q_ 2} I[0, 3, 2, 0] + 
    e_b \Big (3 m_ 0^2 \big (3 I[0, 2, 1, 0] - 9 I[0, 2, 1, 1] + 
          9 I[0, 2, 1, 2] - 3 I[0, 2, 1, 3]
          \nonumber\\
        &+ 2 I[0, 2, 2, 0] - 
          4 I[0, 2, 2, 1] + 2 I[0, 2, 2, 2] - 13 I[0, 2, 3, 0] - 
          13 I[0, 2, 3, 1] + 8 I[0, 2, 4, 0]\big) - 
       4 \big (3 I[0, 3, 1, 1] 
       \nonumber\\
        &- 9 I[0, 3, 1, 2] + 9 I[0, 3, 1, 3] - 
           3 I[0, 3, 1, 4] + 2 I[0, 3, 2, 1] - 4 I[0, 3, 2, 2] + 
           2 I[0, 3, 2, 3] - 13 (I[0, 3, 3, 1] 
           \nonumber\\
        &- I[0, 3, 3, 2]) + 
           8 I[0, 3, 4, 1]\big)\Big) - 
    72 e_ {q_ 2} \Big (3 I[0, 3, 2, 1] - 3 I[0, 3, 2, 2] + 
       I[0, 3, 2, 3] + 3 I[0, 3, 3, 0] - 6 I[0, 3, 3, 1] 
       \nonumber\\
        &+ 
       3 I[0, 3, 3, 2] - 3 I[0, 3, 4, 0] + 3 I[0, 3, 4, 1] + 
       I[0, 3, 5, 0]\Big) + 
    e_ {q_ 1}\Big (-4 (4 I_ 3[\mathcal {S}] + 
           3 I_ 3[\mathcal {\tilde S}]) I[0, 3, 4, 0]
           \nonumber\\
        &+ 
        3  \big (I[0, 3, 2, 0] - 3 I[0, 3, 2, 1] + 3 I[0, 3, 2, 2] - 
            I[0, 3, 2, 3] - 3 I[0, 3, 3, 0] - 6 I[0, 3, 3, 1] + 
            3 I[0, 3, 3, 2] 
            \nonumber\\
        &- 3 I[0, 3, 4, 0] + 3 I[0, 3, 4, 1] - 
            I[0, 3, 5, 0]\big)\mathcal {A}[u_ 0] - 
        3 I_ 3[\mathcal {S}] I[0, 3, 5, 0] + 
        3 \chi \big (-I[0, 4, 2, 1] + 3 I[0, 4, 2, 2] 
        \nonumber\\
        &- 
            3 I[0, 4, 2, 3] + I[0, 4, 2, 4] + 3 I[0, 4, 3, 1] - 
            6 I[0, 4, 3, 2] + 3 I[0, 4, 3, 3] - 3 I[0, 4, 4, 1] + 
            3 I[0, 4, 4, 2]
            \nonumber\\
        &+ I[0, 4, 5, 1]\big) \varphi_ {\gamma}[
           u_ 0]\Big)\Bigg]
   \end{align}
   \begin{align}
    &+\frac {m_b \langle \bar q_2 q_2 \rangle} {196608 \pi^3} \Bigg[-3  e_b m_ 0^2 (I[0, 2, 2, 0] - 
       2 I[0, 2, 2, 1] + I[0, 2, 2, 2] - 2 I[0, 2, 3, 0] + 
       2 I[0, 2, 3, 1] + I[0, 2, 4, 0]) 
       \nonumber\\
        &+ 
    4  e_b \Big (I[0, 3, 2, 1] - 2 I[0, 3, 2, 2] + I[0, 3, 2, 3] - 
       2 I[0, 3, 3, 1] + 2 I[0, 3, 3, 2] + I[0, 3, 4, 1]\Big)
       \nonumber\\
        &+ 
    e_ {q_ 2}\big (I_ 3[\mathcal {S}] + 
        I_ 3[\mathcal {\tilde S}]\big) \big (-4  I[0, 3, 4, 0] + 
        3 I[0, 3, 5, 0]\big)\Bigg] \nonumber\\
        &
        +\frac {m_b^2 } {786432 \pi^5}\Bigg[
   e_b \Big (I[0, 4, 1, 3] - 2 I[0, 4, 1, 4] + I[0, 4, 1, 5] - 
       2 I[0, 4, 2, 3] + 2 I[0, 4, 2, 4] + I[0, 4, 3, 3]\Big) \nonumber\\
        &- 
    8 (e_ {q_ 1} - 
       e_ {q_ 2}) f_ {3\gamma} \pi^2 \Big (I[0, 3, 2, 1] - 
        2 I[0, 3, 2, 2] + I[0, 3, 2, 3] - 2 I[0, 3, 3, 1] + 
        2 I[0, 3, 3, 2] + I[0, 3, 4, 1]\Big) \psi^a[u_ 0]\Bigg] \nonumber\\
        &
        +\frac {f_ {3\gamma}} {1048576 \pi^3} \Bigg[e_ {q_ 1} \
I_ 1[\mathcal {V}] (-4  I[0, 4, 5, 0] + 3 I[0, 4, 6, 0]) - 
   6 (e_ {q_ 1} + 11 e_ {q_ 2}) \Big (I[0, 4, 3, 0] - 
       3 I[0, 4, 3, 1] + 3 I[0, 4, 3, 2] 
       \nonumber\\
        &- I[0, 4, 3, 3] - 
       3 I[0, 4, 4, 0] - 6 I[0, 4, 4, 1] + 3 I[0, 4, 4, 2] - 
       3 I[0, 4, 5, 0] + 3 I[0, 4, 5, 1] - I[0, 4, 6, 0]\Big) \psi^
      a[u_ 0]\Bigg]
      \nonumber\\
        &
        +\frac {m_b} {524288 \pi^5}\Bigg[
   e_b \Bigg (-m_ {q_ 2} \Big (I[0, 4, 2, 2] - 2 I[0, 4, 2, 3] + 
          I[0, 4, 2, 4] - 2 I[0, 4, 3, 2] + 2 I[0, 4, 3, 3] + 
          I[0, 4, 4, 2]\Big) 
          \nonumber\\
        &+ 
       m_ {q_ 1} \Big (3 I[0, 4, 1, 2] - 9 I[0, 4, 1, 3] + 
           9 I[0, 4, 1, 4] - 3 I[0, 4, 1, 5] + 2 I[0, 4, 2, 2] - 
           4 I[0, 4, 2, 3] + 2 I[0, 4, 2, 4] 
           \nonumber\\
        &- 
           13 (I[0, 4, 3, 2] - I[0, 4, 3, 3]) + 
           8 I[0, 4, 4, 2]\Big)\Bigg) + 
    36 e_ {q_ 1} m_ {q_ 2}\Big (-I[0, 4, 2, 1] + 3 I[0, 4, 2, 2] - 
       3 I[0, 4, 2, 3] + I[0, 4, 2, 4] 
       \nonumber\\
        &+ 
       3 (I[0, 4, 3, 1] - 2 I[0, 4, 3, 2] + I[0, 4, 3, 3] - 
          I[0, 4, 4, 1] + I[0, 4, 4, 2]) + I[0, 4, 5, 1]\Big) 
          \nonumber\\
        &- 
    16 (5 e_ {q_ 2} m_ {q_ 1} + 
       e_ {q_ 1} m_ {q_ 2}) f_ {3\gamma}\pi^2 \Big (I[0, 3, 2, 0] - 
        3 I[0, 3, 2, 1] + 3 I[0, 3, 2, 2] - I[0, 3, 2, 3] - 
        3 (I[0, 3, 3, 0] - 2 I[0, 3, 3, 1]
        \nonumber\\
        &+ I[0, 3, 3, 2] - 
           I[0, 3, 4, 0] + I[0, 3, 4, 1]) - I[0, 3, 5, 0]\Big) \psi^
       a[u_ 0]\Bigg]
       \nonumber\\
        &
        +\frac{3} {5242880 \pi^5} \Bigg[-36 e_ {q_ 2} I[0, 5, 3, 1] + 
    11  e_b \Big (I[0, 5, 2, 2] - 3 I[0, 5, 2, 3] + 3 I[0, 5, 2, 4] - 
       I[0, 5, 2, 5] - 
       3 (I[0, 5, 3, 2]
       \nonumber\\
        &- 2 I[0, 5, 3, 3] + I[0, 5, 3, 4] - 
          I[0, 5, 4, 2] + I[0, 5, 4, 3]) - I[0, 5, 5, 2]\Big) + 
    36 e_ {q_ 2}\Big (3 I[0, 5, 3, 2] - 3 I[0, 5, 3, 3] + 
        I[0, 5, 3, 4] 
        \nonumber\\
        &+ 
        3 (I[0, 5, 4, 1] - 2 I[0, 5, 4, 2] + I[0, 5, 4, 3] - 
           I[0, 5, 5, 1] + I[0, 5, 5, 2]) + I[0, 5, 6, 1]\Big)\Bigg],
           \end{align}
where $u_0= \frac{M_1^2}{M_1^2+M_2^2} $, and  $ \frac{1}{M^2}= \frac{1}{M_1^2}+\frac{1}{M_2^2}$ with $ M_1^2 $ and $ M_2^2 $ as the Borel parameters in the initial and final states, respectively. Here $e_{q_1(q_2)}$, $m_{q_1(q_2)}$,  and $\langle \bar q_{1(2)} q_{1(2)} \rangle $ are  the electric charge, mass, and condensates of the corresponding light-quark, respectively.   For simplicity we do not present the terms proportional to many higher dimensional operators; however in the numerical computations we take these terms into account. 
  
The functions~$I[n,m,l,k]$, $I_1[\mathcal{F}]$,~$I_2[\mathcal{F}]$,~$I_3[\mathcal{F}]$, and~$I_4[\mathcal{F}]$ are
defined as:
\begin{align}
 I[n,m,l,k]&= \int_{4m_b^2}^{s_0} ds \int_{0}^1 dt \int_{0}^1 dw~ e^{-s/M^2}~
 s^n\,(s-4m_b^2)^m\,t^l\,w^k,\nonumber\\
 I_1[\mathcal{F}]&=\int D_{\alpha_i} \int_0^1 dv~ \mathcal{F}(\alpha_{\bar q},\alpha_q,\alpha_g)
 \delta'(\alpha_ q +\bar v \alpha_g-u_0),\nonumber\\
  I_2[\mathcal{F}]&=\int D_{\alpha_i} \int_0^1 dv~ \mathcal{F}(\alpha_{\bar q},\alpha_q,\alpha_g)
 \delta'(\alpha_{\bar q}+ v \alpha_g-u_0),\nonumber\\
   I_3[\mathcal{F}]&=\int D_{\alpha_i} \int_0^1 dv~ \mathcal{F}(\alpha_{\bar q},\alpha_q,\alpha_g)
 \delta(\alpha_ q +\bar v \alpha_g-u_0),\nonumber
  \end{align}
 \begin{align}
     I_4[\mathcal{F}]&=\int D_{\alpha_i} \int_0^1 dv~ \mathcal{F}(\alpha_{\bar q},\alpha_q,\alpha_g)
 \delta(\alpha_{\bar q}+ v \alpha_g-u_0),\nonumber
 \end{align}
 where $\mathcal{F}$ denotes the corresponding photon DAs.

 \section*{Appendix B: Distribution Amplitudes of the photon }
In Appendix B, the matrix elements $\langle \gamma(q)\vel \bar{q}(x) \Gamma_i q(0) \ver 0\rangle$  
and $\langle \gamma(q)\vel \bar{q}(x) \Gamma_i G_{\mu\nu}q(0) \ver 0\rangle$ associated with the photon DAs are presented as follows \cite{Ball:2002ps}:
\begin{eqnarray*}
\label{esbs14}
&&\langle \gamma(q) \vert  \bar q(x) \gamma_\mu q(0) \vert 0 \rangle
= e_q f_{3 \gamma} \left(\varepsilon_\mu - q_\mu \frac{\varepsilon
x}{q x} \right) \int_0^1 du e^{i \bar u q x} \psi^v(u)
\nonumber \\
&&\langle \gamma(q) \vert \bar q(x) \gamma_\mu \gamma_5 q(0) \vert 0
\rangle  = - \frac{1}{4} e_q f_{3 \gamma} \epsilon_{\mu \nu \alpha
\beta } \varepsilon^\nu q^\alpha x^\beta \int_0^1 du e^{i \bar u q
x} \psi^a(u)
\nonumber \\
&&\langle \gamma(q) \vert  \bar q(x) \sigma_{\mu \nu} q(0) \vert  0
\rangle  = -i e_q \langle \bar q q \rangle (\varepsilon_\mu q_\nu - \varepsilon_\nu
q_\mu) \int_0^1 du e^{i \bar u qx} \left(\chi \varphi_\gamma(u) +
\frac{x^2}{16} \mathbb{A}  (u) \right) \nonumber \\ 
&&-\frac{i}{2(qx)}  e_q \bar qq \left[x_\nu \left(\varepsilon_\mu - q_\mu
\frac{\varepsilon x}{qx}\right) - x_\mu \left(\varepsilon_\nu -
q_\nu \frac{\varepsilon x}{q x}\right) \right] \int_0^1 du e^{i \bar
u q x} h_\gamma(u)
\nonumber \\
&&\langle \gamma(q) | \bar q(x) g_s G_{\mu \nu} (v x) q(0) \vert 0
\rangle = -i e_q \langle \bar q q \rangle \left(\varepsilon_\mu q_\nu - \varepsilon_\nu
q_\mu \right) \int {\cal D}\alpha_i e^{i (\alpha_{\bar q} + v
\alpha_g) q x} {\cal S}(\alpha_i)
\nonumber \\
&&\langle \gamma(q) | \bar q(x) g_s \tilde G_{\mu \nu}(v
x) i \gamma_5  q(0) \vert 0 \rangle = -i e_q \langle \bar q q \rangle \left(\varepsilon_\mu q_\nu -
\varepsilon_\nu q_\mu \right) \int {\cal D}\alpha_i e^{i
(\alpha_{\bar q} + v \alpha_g) q x} \tilde {\cal S}(\alpha_i)
\nonumber \\
&&\langle \gamma(q) \vert \bar q(x) g_s \tilde G_{\mu \nu}(v x)
\gamma_\alpha \gamma_5 q(0) \vert 0 \rangle = e_q f_{3 \gamma}
q_\alpha (\varepsilon_\mu q_\nu - \varepsilon_\nu q_\mu) \int {\cal
D}\alpha_i e^{i (\alpha_{\bar q} + v \alpha_g) q x} {\cal
A}(\alpha_i)
\nonumber \\
&&\langle \gamma(q) \vert \bar q(x) g_s G_{\mu \nu}(v x) i
\gamma_\alpha q(0) \vert 0 \rangle = e_q f_{3 \gamma} q_\alpha
(\varepsilon_\mu q_\nu - \varepsilon_\nu q_\mu) \int {\cal
D}\alpha_i e^{i (\alpha_{\bar q} + v \alpha_g) q x} {\cal
V}(\alpha_i) \nonumber\\
&& \langle \gamma(q) \vert \bar q(x)
\sigma_{\alpha \beta} g_s G_{\mu \nu}(v x) q(0) \vert 0 \rangle  =
e_q \langle \bar q q \rangle \left\{
        \left[\left(\varepsilon_\mu - q_\mu \frac{\varepsilon x}{q x}\right)\left(g_{\alpha \nu} -
        \frac{1}{qx} (q_\alpha x_\nu + q_\nu x_\alpha)\right) \right. \right. q_\beta
\nonumber \\
 && -
         \left(\varepsilon_\mu - q_\mu \frac{\varepsilon x}{q x}\right)\left(g_{\beta \nu} -
        \frac{1}{qx} (q_\beta x_\nu + q_\nu x_\beta)\right) q_\alpha
-
         \left(\varepsilon_\nu - q_\nu \frac{\varepsilon x}{q x}\right)\left(g_{\alpha \mu} -
        \frac{1}{qx} (q_\alpha x_\mu + q_\mu x_\alpha)\right) q_\beta
\nonumber \\
 &&+
         \left. \left(\varepsilon_\nu - q_\nu \frac{\varepsilon x}{q.x}\right)\left( g_{\beta \mu} -
        \frac{1}{qx} (q_\beta x_\mu + q_\mu x_\beta)\right) q_\alpha \right]
   \int {\cal D}\alpha_i e^{i (\alpha_{\bar q} + v \alpha_g) qx} {\cal T}_1(\alpha_i)
\nonumber \\
 &&+
        \left[\left(\varepsilon_\alpha - q_\alpha \frac{\varepsilon x}{qx}\right)
        \left(g_{\mu \beta} - \frac{1}{qx}(q_\mu x_\beta + q_\beta x_\mu)\right) \right. q_\nu
\nonumber \\ &&-
         \left(\varepsilon_\alpha - q_\alpha \frac{\varepsilon x}{qx}\right)
        \left(g_{\nu \beta} - \frac{1}{qx}(q_\nu x_\beta + q_\beta x_\nu)\right)  q_\mu
\nonumber \\ && -
         \left(\varepsilon_\beta - q_\beta \frac{\varepsilon x}{qx}\right)
        \left(g_{\mu \alpha} - \frac{1}{qx}(q_\mu x_\alpha + q_\alpha x_\mu)\right) q_\nu
\nonumber \\ &&+
         \left. \left(\varepsilon_\beta - q_\beta \frac{\varepsilon x}{qx}\right)
        \left(g_{\nu \alpha} - \frac{1}{qx}(q_\nu x_\alpha + q_\alpha x_\nu) \right) q_\mu
        \right]      
    \int {\cal D} \alpha_i e^{i (\alpha_{\bar q} + v \alpha_g) qx} {\cal T}_2(\alpha_i)
\nonumber \\
&&+\frac{1}{qx} (q_\mu x_\nu - q_\nu x_\mu)
        (\varepsilon_\alpha q_\beta - \varepsilon_\beta q_\alpha)
    \int {\cal D} \alpha_i e^{i (\alpha_{\bar q} + v \alpha_g) qx} {\cal T}_3(\alpha_i)
\nonumber \\ &&+
        \left. \frac{1}{qx} (q_\alpha x_\beta - q_\beta x_\alpha)
        (\varepsilon_\mu q_\nu - \varepsilon_\nu q_\mu)
    \int {\cal D} \alpha_i e^{i (\alpha_{\bar q} + v \alpha_g) qx} {\cal T}_4(\alpha_i)
                        \right\}~,
\end{eqnarray*}
where $\varphi_\gamma(u)$ is the DA of leading twist-2, $\psi^v(u)$,
$\psi^a(u)$, ${\cal A}(\alpha_i)$ and ${\cal V}(\alpha_i)$, are the twist-3 amplitudes, and
$h_\gamma(u)$, $\mathbb{A}(u)$, ${\cal S}(\alpha_i)$, ${\cal{\tilde S}}(\alpha_i)$, ${\cal T}_1(\alpha_i)$, ${\cal T}_2(\alpha_i)$, ${\cal T}_3(\alpha_i)$ 
and ${\cal T}_4(\alpha_i)$ are the
twist-4 photon DAs.
The measure ${\cal D} \alpha_i$ is defined as
\begin{eqnarray*}
\label{nolabel05}
\int {\cal D} \alpha_i = \int_0^1 d \alpha_{\bar q} \int_0^1 d
\alpha_q \int_0^1 d \alpha_g \delta(1-\alpha_{\bar
q}-\alpha_q-\alpha_g)~.\nonumber
\end{eqnarray*}

The expressions of the DAs that are entered into the matrix elements above are  as follows:
\begin{eqnarray}
\varphi_\gamma(u) &=& 6 u \bar u \left( 1 + \varphi_2(\mu)
C_2^{\frac{3}{2}}(u - \bar u) \right),
\nonumber \\
\psi^v(u) &=& 3 \left(3 (2 u - 1)^2 -1 \right)+\frac{3}{64} \left(15
w^V_\gamma - 5 w^A_\gamma\right)
                        \left(3 - 30 (2 u - 1)^2 + 35 (2 u -1)^4
                        \right),
\nonumber \\
\psi^a(u) &=& \left(1- (2 u -1)^2\right)\left(5 (2 u -1)^2 -1\right)
\frac{5}{2}
    \left(1 + \frac{9}{16} w^V_\gamma - \frac{3}{16} w^A_\gamma
    \right),
\nonumber \\
h_\gamma(u) &=& - 10 \left(1 + 2 \kappa^+\right) C_2^{\frac{1}{2}}(u
- \bar u),
\nonumber \\
\mathbb{A}(u) &=& 40 u^2 \bar u^2 \left(3 \kappa - \kappa^+
+1\right)  +
        8 (\zeta_2^+ - 3 \zeta_2) \left[u \bar u (2 + 13 u \bar u) \right.
\nonumber \\ && + \left.
                2 u^3 (10 -15 u + 6 u^2) \ln(u) + 2 \bar u^3 (10 - 15 \bar u + 6 \bar u^2)
        \ln(\bar u) \right],
\nonumber \\
{\cal A}(\alpha_i) &=& 360 \alpha_q \alpha_{\bar q} \alpha_g^2
        \left(1 + w^A_\gamma \frac{1}{2} (7 \alpha_g - 3)\right),
\nonumber \\
{\cal V}(\alpha_i) &=& 540 w^V_\gamma (\alpha_q - \alpha_{\bar q})
\alpha_q \alpha_{\bar q}
                \alpha_g^2,
\nonumber \\
{\cal T}_1(\alpha_i) &=& -120 (3 \zeta_2 + \zeta_2^+)(\alpha_{\bar
q} - \alpha_q)
        \alpha_{\bar q} \alpha_q \alpha_g,
\nonumber \\
{\cal T}_2(\alpha_i) &=& 30 \alpha_g^2 (\alpha_{\bar q} - \alpha_q)
    \left((\kappa - \kappa^+) + (\zeta_1 - \zeta_1^+)(1 - 2\alpha_g) +
    \zeta_2 (3 - 4 \alpha_g)\right),
\nonumber \\
{\cal T}_3(\alpha_i) &=& - 120 (3 \zeta_2 - \zeta_2^+)(\alpha_{\bar
q} -\alpha_q)
        \alpha_{\bar q} \alpha_q \alpha_g,
\nonumber \\
{\cal T}_4(\alpha_i) &=& 30 \alpha_g^2 (\alpha_{\bar q} - \alpha_q)
    \left((\kappa + \kappa^+) + (\zeta_1 + \zeta_1^+)(1 - 2\alpha_g) +
    \zeta_2 (3 - 4 \alpha_g)\right),\nonumber \\
{\cal S}(\alpha_i) &=& 30\alpha_g^2\{(\kappa +
\kappa^+)(1-\alpha_g)+(\zeta_1 + \zeta_1^+)(1 - \alpha_g)(1 -
2\alpha_g)\nonumber +\zeta_2[3 (\alpha_{\bar q} - \alpha_q)^2-\alpha_g(1 - \alpha_g)]\},\nonumber \\
\tilde {\cal S}(\alpha_i) &=&-30\alpha_g^2\{(\kappa -\kappa^+)(1-\alpha_g)+(\zeta_1 - \zeta_1^+)(1 - \alpha_g)(1 -
2\alpha_g)\nonumber +\zeta_2 [3 (\alpha_{\bar q} -\alpha_q)^2-\alpha_g(1 - \alpha_g)]\}.
\end{eqnarray}

The numerical values of the parameters used in the DAs are: $\varphi_2(1~GeV) = 0$, 
$w^V_\gamma = 3.8 \pm 1.8$, $w^A_\gamma = -2.1 \pm 1.0$, $\kappa = 0.2$, $\kappa^+ = 0$, $\zeta_1 = 0.4$, and $\zeta_2 = 0.3$.
 
 \end{widetext}

 \bibliography{TbbMolecule}

\end{document}